\documentclass[conference,a4paper]{IEEEtran}
\IEEEoverridecommandlockouts
\usepackage[utf8]{inputenc}
\usepackage[T1]{fontenc}

\usepackage{amsmath,verbatim}
\usepackage{amssymb}
\usepackage{amsthm}
\usepackage[normalem]{ulem}

\usepackage{array}
\usepackage{url}
\usepackage{xcolor}
\usepackage{pgfplots}
\usepackage{tikz}
\usetikzlibrary{calc,decorations}

\newif\ifcomment
\commentfalse

\newcommand{\aw}[1]{\ifcomment {\color{blue}[AW: #1]} \fi}
\newcommand{\ch}[1]{\ifcomment {\color{magenta}[CH: #1]} \fi}
\newcommand{\lh}[1]{\ifcomment {\color{teal}[LH: #1]} \fi}

\newif\ifcommentLater
\commentLaterfalse
\newcommand{\todoLater}[1]{\ifcommentLater {#1} \fi}

\newcommand{\code}{\ensuremath{\mathcal{C}}}
\DeclareMathOperator{\rank}{rk}

\newcommand{\F}{\ensuremath{\mathbb{F}}}
\newcommand{\Fqs}{\ensuremath{\mathbb{F}_{q^s}}}
\newcommand{\Fq}{\ensuremath{\mathbb{F}_{q}}}

\newcommand{\E}{\ensuremath{\mathcal{E}}}
\newcommand{\I}{\ensuremath{\mathcal{I}}}

\renewcommand{\S}{\ensuremath{\mathcal{S}}}
\renewcommand{\P}{\ensuremath{\mathcal{P}}}
\newcommand{\W}{\ensuremath{\mathcal{W}}}
\newcommand{\V}{\ensuremath{\mathcal{V}}}

\newcommand{\psicut}[2]{\psi_{#1}^{#2}}

\newtheorem{definition}{Definition}
\newtheorem{lemma}{Lemma}
\newtheorem{theorem}{Theorem}

\newtheorem{corollary}{Corollary}
\newtheorem{problem}{Problem}
\newtheorem{remark}{Remark}

\DeclareMathOperator{\rk}{rk}
\DeclareMathOperator{\unif}{unif}

\newcommand{\lrangle}[1]{\left\langle #1 \right\rangle}
\newcommand{\qbinom}[3]{\left[ \begin{matrix} #1 \\ #2 \end{matrix} \right]_{#3}}
\newcolumntype{C}{>{$}c<{$}} 

\usepackage{xcolor}
\definecolor{TUMBlau}{RGB}{0,101,189} 
\definecolor{TUMBlauDunkel}{RGB}{0,82,147} 
\definecolor{TUMBlauHell}{RGB}{152,198,234} 
\definecolor{TUMBlauMittel}{RGB}{100,160,200} 

\definecolor{TUMElfenbein}{RGB}{218,215,203} 
\definecolor{TUMGruen}{RGB}{162,173,0} 
\definecolor{TUMOrange}{RGB}{227,114,34} 
\definecolor{TUMGrau}{gray}{0.6} 

\definecolor{TUMGruenHell}{RGB}{0,124,48}
\definecolor{TUMRot}{RGB}{196,7,27}

\usepackage{flushend}
\flushend

\begin{document}

\title{Computational Code-Based Single-Server Private Information Retrieval}
 \author{%
   \IEEEauthorblockN{Lukas Holzbaur\IEEEauthorrefmark{1}, Camilla Hollanti\IEEEauthorrefmark{2}, Antonia Wachter-Zeh\IEEEauthorrefmark{1} \thanks{The work of L. Holzbaur was supported by the Technical University of Munich -- Institute for Advanced Study, funded by the German Excellence Initiative and European Union 7th Framework Programme under Grant Agreement No. 291763 and the German Research Foundation (Deutsche Forschungsgemeinschaft, DFG) under Grant No. WA3907/1-1. \newline The work of C. Hollanti was supported by the Academy of Finland, under Grant No. 303819 and by the Technical University of Munich -- Institute for Advanced Study, funded by the German Excellence Initiative and the EU 7th Framework Programme under Grant Agreement No. 291763, via a Hans Fischer Fellowship.}}
   \IEEEauthorblockA{\IEEEauthorrefmark{1}Institute for Communications Engineering, Technical University of Munich, Germany\\
     Emails: \{lukas.holzbaur, antonia.wachter-zeh\}@tum.de}
   \IEEEauthorblockA{\IEEEauthorrefmark{2}Department of Mathematics and Systems Analysis, Aalto University, Finland\\
     Email: camilla.hollanti@aalto.fi}
 }

\maketitle

\begin{abstract}
A new computational private information retrieval (PIR) scheme based on random linear codes is presented. 
A matrix of messages from a McEliece scheme is used to query the server with carefully chosen errors. The server responds with the sum of the scalar multiple of the rows of the query matrix and the files. The user recovers the desired file by erasure decoding the response. Contrary to code-based cryptographic systems, the scheme presented here enables to use truly random codes, not only codes disguised as such. Further, we show the relation to the so-called \emph{error subspace search problem} and \emph{quotient error search problem}, which we assume to be difficult, and show that the scheme is secure against attacks based on solving these problems. 
\end{abstract}

\section{Introduction}

Private information retrieval (PIR) was first introduced in~\cite{chor1995private}, enabling a user to retrieve a data item from a database without revealing the identity of the retrieved item to the system owner. A trivial solution would be to download the whole database, which is also the possibility to achieve information theoretic privacy with a single server. 
This solution is infeasible for modern storage systems that can contain a huge number of potentially big files. One possible solution to achieve 
better retrieval rates is to replicate the files on several non-colluding servers, allowing for information theoretic privacy, see, \emph{e.g.}, \cite{dvir20162,beimel2002breaking} for early works and  \cite{sun2018capacity,sun2017capacity,banawan2018capacity,freij2017private} for more recent literature\footnote{Note that in the earlier works, the PIR rate (or its inverse) is referred to as communication complexity, and takes into account both upload and download cost. More recent works typically ignore the upload cost, assuming that the query size is negligible compared to the file size.}. While this allows for schemes of higher rate and lower computational complexity, the assumption of no collusion between some or all of the servers is regarded as unpractical in many use cases.

Schemes for single-server computational PIR have been presented in \cite{kushilevitz1997replication,lipmaa2005oblivious,gentry2005single}, relying on ``pre-quantum'' cryptographic paradigms, \emph{i.e.}, on paradigms that will be rendered insecure once a sufficiently powerful quantum computer exists.

The practicality of the existing computational PIR schemes was  discussed in  \cite{sion2007computational}, concluding that in a realistic setting, the execution of these schemes would take more time than the trivial solution of downloading the whole database due to the computational complexity on the server side. \todoLater{\ch{This is probably not true if we assume big files. For big files, I guess they can use the block retrieval and will be better. But I would like to avoid the term PBR.}\lh{Isn't it? The problem that very complex operations have to be performed for every bit of the file remains the same.}} Following this discussion, further effort has been made in finding computational PIR schemes with lower computational complexity.  A  computationally efficient lattice-based computational PIR scheme was proposed in \cite{aguilar2007lattice}, which can plausibly be executed in less time than the trivial solution. A practical attack to this scheme for databases with a small number of elements was found in \cite{liu2016cryptanalysis}. However, this is not a very big drawback, as modern databases and storage systems tend to contain a large number of files.

In  \cite{gentry2009fully}, the first fully homomorphic encryption (FHE) scheme  was constructed using lattice-based cryptography. Following this breakthrough, \cite{yi2012single} gave a general construction from a FHE scheme to PIR. 
Furthermore, they give an instance of this construction which is practical and outperforms the scheme in \cite{aguilar2007lattice}. Other PIR schemes based on homomorphic encryption were proposed recently in \cite{kiayias2015optimal,aguilar2016xpir,lipmaa2017simpler,gentry2019compressible}. Building on the protocol of \cite{aguilar2016xpir}, a method to significantly decrease the query size was introduced in \cite{angel2018pir}.

This paper is the first to provide a computational PIR scheme based on codes, and can be seen as a counter-part to the lattice-based scheme of \cite{aguilar2007lattice} along the same lines as code-based and lattice-based cryptography are connected in general. 
The query to the sever can be considered as a matrix whose rows contain corrupted codewords of a secret code.
The server then responds with the scalar product of the query matrix and the files and the user can recover the requested file by erasure decoding. Depending on the parameters, the achieved PIR rates are comparable to the existing computational PIR schemes of \cite{yi2012single,aguilar2007lattice}. The complexity, which is the bottleneck of current computational schemes, benefits from all calculations being over binary extension fields, which is advantageous for implementation.
\begin{remark}
This computational PIR scheme has recently been broken for all relevant parameters. For details see \cite{bordage2020privacy}.
\end{remark}
\section{Notation}
Let $F_{q}$ denote the finite field of order $q$ and $\F_{q^s}$ its extension field of extension degree $s$.
We write $[a,b]$ for the set $\{a,a+1,...,b\}$ and if $a=1$ we write $[b]$.
We denote a linear code $\code$ over $\F_{q}$ of length $n$ and dimension $k$ by $[n,k]_{q}$. Let $G$ be a generator matrix of $\code$. We say a set $\mathcal{I}\subseteq [n]$ with $|\mathcal{I}| \geq k$ is an information set of the code if $\rank(G|_{\I}) = k$, where $G|_{\I}$ denotes the restriction of the matrix $G$ to the columns in indexes by~$\mathcal{I}$.

\section{Preliminaries}
We begin by defining some basic functions required for the description of the PIR scheme.

\begin{definition} \label{lem:phimap}
Let $\E \subseteq [n]$ and $E \in \F^{\xi \times |\E|}$. Denote by $M_{\E}$ the $n\times n$ identity matrix with all rows index by $[n]\backslash \E$ deleted. The map $\phi(E,\E)$ is given by
  $\phi_n(E,\E) = E \cdot M_{\E} .$
\end{definition}

For example, consider the mapping
\begin{align*}
    \phi_n\left(\left(\begin{smallmatrix}
    1&2\\
    3&4
    \end{smallmatrix}\right),\{2,4\}\right) = 
    \left(\begin{smallmatrix}
    1&2\\
    3&4
    \end{smallmatrix}\right) \cdot \hspace{-35pt}
    \underbrace{\left(\begin{smallmatrix}
    0&1&0&0\\
    0&0&0&1
    \end{smallmatrix}\right)}_{\text{identity matrix rows $1$ and $3$ deleted}} \hspace{-35pt}=
    \left(\begin{smallmatrix}
    0&1&0&3\\
    0&2&0&4
    \end{smallmatrix}\right) \ .
\end{align*}

In the following we need to be able to ``cut" out the part of an element contained in a certain subspace. 
\begin{definition}\label{def:cutSubspace}
Let $\Gamma = \{\gamma_1,\gamma_2,...,\gamma_s\}$ be a basis of $\F_{q^s}$ over $\F_q$ and $\alpha$ be an element $\alpha \in \F_{q^s}$ with $\alpha = \sum_{j=1}^{s} \alpha_j \gamma_j , \alpha_i \in \F_{q}$. For a subspace $W$ with basis $\W$ such that $\W \subset \Gamma$ we define
\begin{equation*}
    \psicut{\Gamma}{W}(\alpha) = \sum_{\gamma_j \in \W} \alpha_j \gamma_j \ .
\end{equation*}
\end{definition}
Note that for any element $\alpha \in W$ of a subspace of $\Fqs$ and element $\alpha' \in \Fqs / W$ of the quotient space it holds that
\begin{equation*}
    \psicut{\Gamma}{W}(\beta \alpha + \beta' \alpha' ) = \beta \alpha \  \forall \ \beta,\beta' \in \Fq \ .
\end{equation*}

\section{A Code-Based Computational PIR scheme}

In a computational PIR scheme, a user generates a query $Q^i$ from a set of secret information $\S$ and a set of public information $\P$. For each such query the server replies with some $A^i$, which is a function of the received query $Q^i$, the $m$ files $X^1,...,X^m$ stored on the server, and the public information $\mathcal{P}$. The scheme is said to be \emph{correct} if the user can recover the desired file from the replies of the servers.

\subsection{System Model}
We consider a single server storing $m$ files, i.e., in total we store $X \in \F_{q}^{L \times m(s-v)(n-k)}$, where each file  $X^l \in \F_{q}^{L(s-v)(n-k)}$ is given by a submatrix of $(s-v)(n-k)$ columns (compare Figure~\ref{fig:fileMatrix}). We denote $\delta := (s-v)(n-k)$, this parameter can be considered the required level of subpacketization. We assume that the indices of the files are known to the user.

\begin{figure}
  \centering
  \def\x{*1}

\begin{tikzpicture}

  \coordinate (Xnw) at (0\x,1\x);
  \node[draw=none] () at ($(Xnw)+(-0.5\x,-0.4\x)$) {$X = $};

  \draw[draw=black] (Xnw) rectangle ($(Xnw)+(1.2\x,-0.8\x)$) node[pos=0.5] {$X^1$};
  \draw[draw=black] ($(Xnw)+(1.2\x,0\x)$) rectangle ++(1.2\x,-0.8\x) node[pos=0.5] {$X^2$};
  \draw[draw=black] ($(Xnw)+(2.4\x,0\x)$) rectangle ++(1.2\x,-0.8\x) node[pos=0.5] {$X^3$};
  \node[draw=none] at ($(Xnw)+(4.2\x,-0.4\x)$) {$\cdots$};
  \draw[draw=black] ($(Xnw)+(4.8\x,0\x)$) rectangle ++(1.2\x,-0.8\x) node[pos=0.5] {$X^m$};
    
  \draw [decorate,decoration={brace,amplitude=3pt}] ($(Xnw)+(6.1\x,0\x)$) -- ++(0\x,-0.8\x) node [black,midway,xshift=0.3\x cm] {$L$};
  \draw [decorate,decoration={brace,amplitude=3pt}] ($(Xnw)+(1.2\x,0.1\x)$) -- ++(1.2\x,0\x) node [black,midway,yshift=0.4\x cm] {$(s-v)(n-k)$};
  
\end{tikzpicture}

  \caption{Illustration of file matrix $X$.}
  \label{fig:fileMatrix}
  \vspace{-20pt}
\end{figure}

\subsection{Query}

The user chooses a random $[n,k]_{q^s}$ code $\code$. \todoLater{\lh{Not sure if we should restrict to codes without constant positions, i.e., zero columns in the generator matrix. It's a reasonable restriction but might cause problems in the analysis.} \aw{Where is the problem in the analysis?} \lh{In the linear dependency attack we rely on the randomness of $D$. But maybe it's actually not a problem, since we only need it to say something about information sets of the code and all-zero columns are clearly not part of those. I'll check.}} Let $D \in \F_{q^s}^{m\delta \times n}$ be a matrix where each row $D_{l,:}$ is chosen uniformly at random from $\code$. Let $\mathcal{I} \subset [n]$ with $|\mathcal{I}|=k$ be a randomly chosen information set of $\code$ and denote its complement by $\E = [n]\setminus \I$.
Further, the user chooses a random basis $\Gamma = \{\gamma_{1},\gamma_{2},...,\gamma_{s}\}$ of $\F_{q^s}$ over $\F_{q}$. We denote by $V$ the $\F_q$-linear subspace of $\F_{q^s}$ of dimension $v$ spanned by $\V=\{\gamma_{1},...,\gamma_{v}\}$, where $v < s$, and by $W$ be the $s-v$ dimensional subspace spanned by $\W = \{\gamma_{v+1},...,\gamma_{s}\}$, i.e., the quotient space $\F_{q^s} / V$. The user chooses a matrix $\hat{E}\in V^{m\delta \times n-k}$ i.i.d. at random. We denote $E=\phi_n(\hat{E},\E)$.

\todoLater{\lh{The division into ``useful'' and ``useless'' subspace doesn't have to be the same in each column and choosing them randomly avoids that an attacker can expect $\F_q$-linear combinations of them to have any specific structure. However, it makes ``guessing'' one of the used subspaces easier, and since that is our main concern currently, I now choose all subspaces the same.}}

Let $\widehat{\Delta} \in W^{(s-v)(n-k)\times n-k}$ be chosen i.i.d. random from matrices of full row-rank over $\F_q$, i.e., with 
$\rk_q(\widehat{\Delta}) = (s-v)(n-k) ,$
and denote $\Delta = \phi(\widehat{\Delta},\E)$.

The query for file $X^i$ is given by
\begin{equation}\label{eq:query}
  Q^i = D + E + \Delta \otimes e_i^m \ ,
\end{equation}
where $e_i^m \in F_2^{m \times 1}$ denotes the $i$-th unit vector and $\otimes$ denotes the Kronecker product. An illustration of the query matrix $Q^i$ is given in Figure~\ref{fig:queryMatrix}.

\begin{figure}
  \centering
  \def\x{*1}

\begin{tikzpicture}
  
  \coordinate (Dnw) at (0\x,3\x);
  \node[draw=none] () at ($(Dnw)+(-0.5\x,-3\x)$) {$Q^i = $};

  \draw[draw=black] (Dnw) rectangle ($(Dnw)+(2\x,-6\x)$) node[pos=0.5] {$D$};
  \node[draw=none] at ($(Dnw)+(2.3\x,-3\x)$) {$+$};

  \draw [decorate,decoration={brace,amplitude=3pt}] ($(Dnw)+(0\x,0.1\x)$) -- ++(2\x,0\x) node [black,midway,yshift=0.3\x cm] {$n$};

  \coordinate (Enw) at ($(Dnw) + (2.6\x,0\x)$);

  \node[draw=none] at ($(Enw)+(2.3\x,-3\x)$) {$+$};

  \draw[draw=none, fill = TUMBlau!30!white] ($(Enw)+(0.0\x,0\x)$) rectangle ++(0.3\x,-6\x);
  \draw[draw=none, fill = TUMBlau!30!white] ($(Enw)+(0.55\x,0\x)$) rectangle ++(0.3\x,-6\x);
  \draw[draw=none, fill = TUMBlau!30!white] ($(Enw)+(1.3\x,0\x)$) rectangle ++(0.3\x,-6\x);

  \draw[draw=black] (Enw) rectangle ($(Enw)+(2\x,-6\x)$) node[pos=0.5] {$E$};
  
  \coordinate (Knw) at ($(Enw) + (2.6\x,0\x)$);

  \draw [decorate,decoration={brace,amplitude=3pt}] ($(Knw)+(2.1\x,0\x)$) -- ++(0\x,-6\x) node [black,midway,xshift=0.4\x cm, rotate=270] {$m(s-v)(n-k)$};

  \coordinate (Deltanw) at ($(Knw) + (0\x,-3.6\x)$);

  \draw[draw=none, fill=TUMGruenHell!30!white, dotted] ($(Deltanw)+(0\x,0\x)$) rectangle ++(0.3\x,-1.2\x);
  \draw[draw=none, fill=TUMGruenHell!30!white,dotted] ($(Deltanw)+(0.55\x,0\x)$) rectangle ++(0.3\x,-1.2\x);
  \draw[draw=none, fill=TUMGruenHell!30!white,,dotted] ($(Deltanw)+(1.3\x,0\x)$) rectangle ++(0.3\x,-1.2\x);

  \draw[draw=black, dashed] (Deltanw) rectangle ($(Deltanw)+(2\x,-1.2\x)$) node[pos=0.5] {$\Delta$};

  \draw[draw=black] (Knw) rectangle ($(Knw)+(2\x,-6\x)$) node[pos=0.5] {$\Delta \otimes e_i^m$};

\end{tikzpicture}

    \vspace{-7pt}
  \caption{Illustration of query matrix $Q^i$.}
  \label{fig:queryMatrix}
  \vspace{-13pt}
\end{figure}

\subsection{Response}
The server receives the query $Q^i \in \F_{q^s}^{m \times n}$ and responds with
\begin{equation}
  \label{eq:response}
  A^i = X \cdot Q^i  \in \F_{q^s}^{L \times n}\ ,
\end{equation}
i.e., with a matrix where each row is an $\F_q$-linear combination of the rows of $Q^i$ with coefficients given by the respective row of $X$.

\subsection{Decoding}
Denote the $j$-th unit vector of length $n$ by $e_j^n$. The user receives a matrix where the $z$-th row is given by
\begin{align*}
  &A^i_{z,:} = X_{z,:} \cdot Q^i \\
&= \left(\sum_{l=1}^m X_{z,:}^l \cdot (D_{(l-1)\delta+1:l\delta,:} + E_{(l-1)\delta+1:l\delta,:}) \right) + X_{z,:}^i\cdot \Delta \\
  &= \underbrace{\left(\sum_{l=1}^m X_{z,:}^l \cdot D_{(l-1)\delta+1:l\delta,:}\right)}_{\in \code} \\
  &\qquad+ \underbrace{\left(\sum_{l=1}^m X_{z,:}^l \cdot E_{(l-1)\delta+1:l\delta,:} \right) + X_{z,:}^i \cdot \Delta}_{\text{zero in positions } [n] \setminus \E} \ .
\end{align*}
As the positions $[n] \setminus \E$ are an information set of $\code$ by definition and the set $\E$ is known to the user, the entire vector $\sum_{l=1}^m X_{z,:}^l \cdot D_{(l-1)\delta+1:l\delta,:}$ can be recovered and thereby
\begin{align*}
  A_{z,:}^i &- \left(\sum_{l=1}^m X_{z,:}^l \cdot D_{(l-1)\delta+1:l\delta,:}\right) \\
  &= \underbrace{\left(\sum_{l=1}^m X_{z,:}^l \cdot E_{(l-1)\delta+1:l\delta,:} \right)}_{\in V^{1\times n}} + \underbrace{X_{z,:}^i\cdot \Delta}_{\in W^{1\times n}} \ .
\end{align*}
Applying the function from Definition~\ref{def:cutSubspace} with respect to $W$ yields
\begin{align*}
  \psicut{\Gamma}{W} \left( \left(\sum_{l=1}^m X_{z,:}^l \cdot E_{(l-1)\delta+1:l\delta,:} \right) + X_{z,:}^i\cdot \Delta \right) = X_{z,:}^i\cdot \Delta\ .
\end{align*}
As $\hat{\Delta}$ is of full row-rank over $\F_q$, so is $\Delta$. Hence, the vector $X_{z,:}^i$ can be recovered and finally the entire file $X^i$ by performing these steps on each row $z\in [L]$.

\subsection{Analysis}
The upload, i.e., the size of the query, in bits is
\begin{equation*}
  H(Q^i) = m\delta n \log_2(q^s) = m\delta ns \log_2(q) .
\end{equation*}
The download, i.e., the size of the response, in bits is
\begin{equation*}
  H(A^i) = Ln \log_2(q^s) = Lns \log_2(q) .
\end{equation*}

\begin{theorem}
  The rate of the scheme is given by
  \begin{align*}
    R_{\mathrm{PIR}} &= \frac{L\delta \log_2(q)}{m\delta ns \log_2(q) + L n s \log_2(q)} \\[0.4ex]
                   &= \frac{L}{m\delta + L}\left( 1-\frac{k+\frac{v}{s}(n-k)}{n} \right). \\
  \end{align*}
\end{theorem}

A common assumption in literature is that the size of the file is much larger than the number of files, i.e., $L >> \delta m$. In this case it is reasonable to neglect the upload cost in the calculation of the rate of the scheme.

\begin{corollary}[PIR Rate]\label{cor:PIRrate}
  For $L >> \delta m$, the rate of the scheme is
  \begin{align*}
    R_{\mathrm{PIR}} &\approx 1-\frac{k+\frac{v}{s}(n-k)}{n} \ .
  \end{align*}
\end{corollary}
\todoLater{\lh{I think this formulation looks quite nice, it's pretty similar to the regular PIR rates. Note that $v=0$ is insecure as we don't add any ``errors'' except in the positions of the file that we want in this case. The first potentially secure case is $v=1$, which would basically mean: everything is an extension field, while the ``errors'' added in the undesired file positions are binary, and the remaining ``error dimensions'' are used to retrieve the file. So the best rate we can get is
  \begin{equation*}
    1-\frac{k+\frac{1}{s}(n-k)}{n} \stackrel{s\rightarrow \infty}{\longrightarrow} 1-\frac{k}{n}\ .
  \end{equation*}
  The actual choice of all parameters for a certain security level is determined by the attacks of Section~\ref{sec:subspaceAttack}.
}}

\section{Security Analysis}

\subsection{Subspace Attack} \label{sec:subspaceAttack}

The security of the system is based on the idea that it is difficult for the attacker to differentiate which rows of $D$, i.e., elements of $\code$, are corrupted by elements from a different subspace than the other rows (indicated by the green columns of $\Delta$ in Figure~\ref{fig:queryMatrix}).

The security of our system is therefore tightly related to the following search problem.
\begin{problem}[Error Subspace Search Problem]\label{prob:error-subspace-search}
Given a set of words in $\Fqs^n$ which are each the sum of a codeword of a random code $\mathcal{C}$ and an error vector. 
Find a $v$-dimensional subspace that contains the largest possible number of these error vectors.
\end{problem}
Solving this general problem efficiently would break our system. 
Since the code of our system is unknown, it appears as a random code to an attacker.
It is known that decoding a random code (i.e., explicitly finding the error vector(s)) is an NP hard problem.
Problem~\ref{prob:error-subspace-search} is easier than decoding as we do not want to decode all words (or many), but find the $v$-dimensional subspace that contains the most error vectors. 
However, we are not aware of how to find this subspace other than just trying all $v$-dimensional subspaces which results in an exponential complexity.
Once this $v$-dimensional subspace is known, an approach to break our system is derived in the following.

The rows of the matrix $Q^T$ are considered as the basis of a code. \todoLater{\lh{Change order of attacks, then we can say that the dimension of the $Q^T$ code is actually $n$ w.h.p.?}} As the rows of $D$ are random elements of a $k$-dimensional vector space, another basis of this code is given by
\begin{equation*}
    \lrangle{Q^T} = \lrangle{\left(D\cdot A , \hat{E} + \hat{\Delta} \otimes e_i^m\right)^T} , 
\end{equation*}
for some full-rank matrix $A\in\F^{n\times k}$. Recall that the elements of $\hat{E}$ are from the space $V$ and the elements of $\hat{\Delta}$ are from the quotient space $W$. It follows that all elements in $(\hat{E} + \hat{\Delta} \otimes e_i^m )^T $ are from $V$, except for the ones corresponding to file $i$, which can be from the entire field $\F_{q^s}$.  Therefore, if the attacker is able to find such a basis, the index of the desired file can easily be determined.
Hence we can restate the problem as: find a subspace of $\lrangle{Q^T}$ such that all positions except for those corresponding to one file are from a subspace of dimension $\dim(V) = v$.

Once a suitable subspace $V$ is known (or for any guessed subspace), an attacker can proceed by the following procedure: 
\begin{enumerate}
\item Consider the $[m\delta,n]$ code spanned by $Q^T$. Puncture the positions belonging to the file $l$.
\item Calculate a parity-check matrix of this code. This matrix spans the dual code of dimension $(m-1)\delta-n$.
\item Extend the parity-check matrix to the subfield. If everything is random, the dimension of the subfield subcode is $\max\{(m-1)\delta- ((m-1)\delta - n)s ,0\}$ w.h.p. As $m>>n$, this is almost certainly $0$.
\item If the dimension of the subcode is zero, then $l \neq i$. If it is non-zero, then $l=i$ w.h.p.
\end{enumerate}
This attack is successful w.h.p. for all parameters that lead to a reasonable rate. However, it requires that the attacker knows that subspace $V$ in order to determine the dimension of the corresponding subcode. Hence, to prevent this attack, the system parameters need to be chosen such that the probability of the attacker guessing the correct subspace is small. 

The number of $v$-dimensional subspaces of an $s$-dimensional space (where $v\leq s$) is given by 
the Gaussian binomial coefficient, i.e., 
\begin{equation*}
  \qbinom{s}{v}{q}  = \frac{(1-q^s)(1-q^{s-1})...(1-q^{s-v+1})}{(1-q^v)(1-q^{v-1})...(1-q)} .
\end{equation*}

Instead of guessing the actual $v$-dimensional subspace $V$, the attacker can also guess a larger subspace in the hope that it contains the correct space $V$, as any subspace subcode can be expected to be empty if the number of files $m$ is large (the probability approaches $1$ as $m \rightarrow \infty$). The probability of picking a space containing $V$ depends on the number of possible extensions spaces,
i.e., the number of higher dimensional subspaces a smaller subspace is contained in.
\begin{lemma}
  Every $v$-dimensional subspace of $\,\F_{q^s}$ is a subspace of
  \begin{equation*}
    \qbinom{s-v}{z-v}{q}
  \end{equation*}
  subspaces of dimension $z$.
\end{lemma}
\begin{IEEEproof}
Let $V$ be any $v$-dimensional subspace of $\F_{q^s}$ and $Z$ be a $z$-dimensional subspace containing it. Then there is a one-to-one mapping between the $Z$ and the $z-v$-dimensional subspaces of the quotient space $\F_{q^s}/ V$.
\end{IEEEproof}
The attack is successful if the attacker picks one of these $(s-1)$-dimensional ``superspaces'', which happens with probability
\begin{align*}
  \Pr&\{V\subseteq Z\} = \underbrace{\qbinom{s-v}{s-1-v}{q}}_{\substack{\text{\# of $(s-1)$-dim.}\\ \text{extension spaces}}} \quad \ \cdot \quad  \underbrace{\qbinom{s}{s-1}{q}^{-1}}_{\substack{\text{inverse of \# of}\\ \text{$(s-1)$-dim. spaces}}} \ ,
\end{align*}
if the space $Z$ is chosen uniformly at random. 
\todoLater{\lh{There could be a better choice for the attacker, \emph{e.g.}, there is no point in trying subspaces with very large intersections.}
\aw{Are you sure that there is no point in trying subspaces with large intersection? Can we \emph{e.g.}, see that we guessed a subspace that intersects with $V$ in $v-1$ dimensions? Compared to not intersecting at all?}

\aw{What about an "iterative" process: Start with $s-1$, find superspace, then search in this $s-1$-dimensional subspace a $s-2$-dimensional superspace and so on until we reach $v$?}
\lh{I'm assuming the worst case here, which is that if the space we choose contains $V$ as a subspace, the attacker is done. So it is not necessary to guess the actual space $V$, but only a superspace. That's what I meant by it being unnecessary to guess spaces with big intersections. Say the attacker has guess the spaces $Z_1$ with $\{V_1,V_2,V_3\}\subset Z_1$ and $Z_2$ with $\{V_4,V_5,V_6\}\subset Z_2$, then there is no point in also trying $Z_3$ with $\{V_1,V_4,V_5\}$, since none of those are the correct one. Of course this is much simplified, but that's what I meant by not trying spaces with large intersections.}}
To prevent the attack, we require the inverse of this probability to be larger than the security level of the scheme.

\subsection{Linear Dependency Attack} \label{sec:linDepAttack}
The goal of the attacker is to determine for which $l\in [m]$ the corresponding rows in $Q$ differ from the other rows. In this section, we discuss an attack that aims at directly finding the file index~$i$ by comparing the probability of rows of the query matrix being independent, given that positions corresponding to $l$ are included or not.
We can therefore say that if one can efficiently solve the following problem, our system would be broken.
\begin{problem}[Quotient Error Search Problem]\label{prob:quotient-error-subspace-search}
Given a set of words in $\Fqs^n$ which are each the sum of a codeword of a random code $\mathcal{C}$ and an error vector from a subspace $\F_{q^v}^{n}$, except for one, to which an additional error vector from the quotient space $\Fqs^n / \F_{q^v}^n$ is added. 
Find the word with the additional error vector from the quotient space.
\end{problem}

\todoLater{\lh{The following is for the special case where $E$ is from a subfield. If $E$ is from a subspace, it gets even more complicated to analyze. It should increase the probability, but that needs to be shown. Also it should be considered what happens if we take small submatrices out of the query matrix. In this case it will depend on the probability of picking some number of "error" positions, just like in McEliece.}
\aw{Is this only for the subfield case? If the subspace case is too complicated we can just describe the subfield case and say that it can be generalized to the subspace case but this is left out to improve readability and due to lack of space...}}
We analyze the probability of a square submatrix of $Q$ being of full rank if it does not contain any rows corresponding to the $i$-th file. This probability differs from the probability for a submatrix containing rows corresponding to the $i$-th file, as the probability of a matrix being full-rank decreases with the size of the subspace. For simplicity we only consider the case where $\F_{q^v}$ is a subfield of $\Fqs$ and leave the generalization to arbitrary subspaces for an extended version of this work.
\begin{theorem}
 Let $v|s$. Then for any $\I \subset [m\delta]\setminus \{(i-1)\delta+1,...,i\delta\}$ with $|\S| = n$ it holds that
\begin{align*}
  \Pr&\{\rk_{q^s}(D_{\I,:} + E_{\I,:}) = n\} \geq \left(\prod_{j=n-k+1}^n 1-\frac{1}{q^{sj}}\right)\\
  &\cdot \left(\prod_{j=1}^{n-k} \left(1-\frac{1}{q^{sj}}\right) -\left(1-\prod_{j=k+1}^n \left(1-\frac{1}{q^{sj}}\right)\right)  \right)
\end{align*}
\end{theorem}
\begin{IEEEproof}
Without loss of generality assume that $\E = [n-k]$. By slight abuse of notation we drop the index $\I$ in the following, i.e., instead of $D_{\I,:}$ and $E_{\I,:}$ we simply write $D$ and $E$.

Let $B\in \F_{q^s}^{n\times n}$ be chosen uniformly at random from all full-rank matrices with
\begin{align}
  B \cdot (D+E) &=
              \left(
  \begin{smallmatrix}
    B_{1} \\
    B_{2}
  \end{smallmatrix} \right)
              \cdot \left(
              \left(
  \begin{smallmatrix}
    D_{1,1} & D_{1,2} \\
    D_{2,1} & D_{2,2}
  \end{smallmatrix} \right)
              + \left(
  \begin{smallmatrix}
    \hat{E}_1 & \mathbf{0}_{k \times k} \\
    \hat{E}_2 & \mathbf{0}_{n-k \times k}
  \end{smallmatrix} \right) \right) \nonumber \\
    &= \left(\begin{smallmatrix}
      D_1' & D_2' \\
      \mathbf{0}_{n-k\times n-k} & \mathbf{0}_{n-k\times k}
    \end{smallmatrix} \right)
  + \left(
  \begin{smallmatrix}
    \hat{E}_1' & \mathbf{0}_{k \times k} \\
    \hat{E}_2' & \mathbf{0}_{n-k \times k}
  \end{smallmatrix} \right) \ . \label{eq:BDE}
\end{align}
Note that such a matrix always exists since the rows of $D$ are taken from a $k$-dimensional subspace and $E$ is only supported on $\E$.

The matrix $B\cdot(D+E)$ is of full rank if and only if $D_2'$ and $\hat{E}_2'$ are of full rank, therefore
\begin{align*}
  &\Pr\{\rank(D+E) = n \} = \Pr\{\rank(D_2') = k \land \rank(E_2') = n-k  \} \\
  &\quad= \Pr\{\rank(D_2') = k\} \cdot \Pr\{\rank(E_2') = n-k | \rank(D_2') = k \} \ .
\end{align*}
Since $[n] \setminus \E$ is an information set of $\code$ by definition, the matrix $D_2'$ is of full rank if and only if the matrix $D$ contains a basis of the code $\code$. Let $G_s$ be a generator matrix of the code $\code$, then there is an $U\in \F_{q^s}^{n\times k}$ such that
\begin{equation*}
  U \cdot G_s = D \ .
\end{equation*}
The codewords in $D$ are chosen uniformly at random, which is equivalent to $U\sim \unif(\F_{q^s}^{n\times k})$. Since the generator matrix $G_s$ is full-rank by definition, the multiplication is rank preserving. Hence, it holds that $\rank(D) = k$, i.e., the matrix $D$ contains a basis of $\code$, if and only if $\rank(U) = k$, which is well-known to be
\begin{align*}
  \Pr\{\rank(D_2') = k\} = \Pr\{ \rank(U) = k\} = \prod_{j=n-k+1}^n 1-\frac{1}{q^{sj}} \ .
\end{align*}
Now consider the bottom part of the matrix. From (\ref{eq:BDE}) we get
\begin{align*}
  \mathbf{0}_{n-k \times n} &=
    B_{2}
              \cdot
  \begin{pmatrix}
    D_{1,1} & D_{1,2} \\
    D_{2,1} & D_{2,2}
  \end{pmatrix}\\
            &=
    B_{2}
              \cdot \left(
  \begin{pmatrix}
    U_{1} \\
    U_{2}
  \end{pmatrix} \cdot G_s \right) \\
            &\!\!\!\!\!\!\Rightarrow
    B_{2}
              \cdot
  \begin{pmatrix}
    U_{1} \\
    U_{2}
  \end{pmatrix} = \mathbf{0}_{n-k \times k} \ .
\end{align*}
Since $\rank(U)=k$ and $\rank(B) = n$ by assumption, it follows that $B_2$ is a basis of the dual space of $U$. As $U$ is chosen uniformly at random, every full rank $U$ is equally likely and therefore also any $B_2$. From (\ref{eq:BDE}) we further get
\begin{equation*}
  \hat{E}_2' =
    B_{2} \cdot
  \begin{pmatrix}
    \hat{E}_1\\
    \hat{E}_2
  \end{pmatrix}  = B_2 \cdot \hat{E} \ ,
\end{equation*}
where $B_{2} \in \F_{q^s}$ and $\hat{E}_1, \hat{E}_2 \in \F_q$. Let $M \sim \unif(\F_{q^s}^{n-k\times n})$. We are interested in the probability
\begin{align*}
  \Pr&\{ \rk(B_2 \cdot \hat{E}) = n-k |\rk(D_2') = k \} \\
     &= \Pr\{ \rk(B_2 \cdot \hat{E}) = n-k |\rk(U) = k \} \\
     &\stackrel{(a)}{=} \Pr\{ \rk(B_2 \cdot \hat{E}) = n-k \} \\
     &= \Pr\{ \rk(M \cdot \hat{E}) = n-k |\rk(M) = n-k \} \\
     &= \frac{\Pr\{ \rk(M \cdot \hat{E}) = n-k \land \rk(M) = n-k \}}{\Pr\{\rk(M) = n-k \}} \\
     &\geq \Pr\{ \rk(M \cdot \hat{E}) = n-k \land \rk(M) = n-k \} \\
     &= 1-\Pr\{ \rk(M \cdot \hat{E}) < n-k \lor \rk(M) < n-k \} \\
     &\geq 1-(\Pr\{ \rk(M \cdot \hat{E}) < n-k\}+ \Pr\{ \rk(M) < n-k\} ) \\
     &= 1\!-\!(1\!-\!\Pr\{ \rk(M \! \cdot \! \hat{E}) \!= \!n-k\})\!- \!\Pr\{ \rk(M)\! < \!n-k\} ) \\
     &= \Pr\{ \rk(M \cdot \hat{E}) = n-k\}- \Pr\{ \rk(M) < n-k\} \ ,
\end{align*}
where $(a)$ holds because $\hat{E}$ is independent of $U$ and $B_2$ is uniformly distributed if $U$ is uniformly distributed over all full rank matrices.
\todoLater{\lh{Maybe this needs to be shown in more detail, it basically says that given a random subspace, every subspace of a given dimension is equally likely to be in it's dual. Also given random subspace, every full rank matrix is equally likely to form a basis of this space or a space containing it.}}
To obtain the first probability, we fix a basis of $\F_{q^s}$ over $\F_q$ and consider the extension of $M \in \F_{q^s}^{n-k \times n}$ to $\bar{M} \in \F_{q}^{s(n-k) \times n}$ obtained by representing every element in this basis. \todoLater{\lh{This is where we need the condition $v|s$, since we can't fix a basis of $\Fqs$ over $\F_{q^v}$ otherwise.}} As $M$ is random over $F_{q^s}$, the matrix $\bar{M}$ is random over $\F_q$. The multiplication of two random matrices is again a random matrix, hence we get $\bar{M} \cdot \hat{E} \sim \unif(\F_{q}^{s(n-k) \times n-k})$ and equivalently, when mapping back to $F_{q^s}$, we get $M \cdot \hat{E} \sim \F_{q^s}^{n-k \times n-k}$. Hence
\begin{equation*}
  \Pr\{ \rk(M \cdot \hat{E}) = n-k\} = \prod_{j=1}^{n-k} 1-\frac{1}{q^{sj}} \ .
\end{equation*}
It follows that
\begin{align*}
  \Pr&\{ \rk(B_2 \cdot \hat{E}) = n-k \} \\
  &\geq \prod_{j=1}^{n-k} \left(1-\frac{1}{q^{sj}}\right) -\left(1-\prod_{j=k+1}^n \left(1-\frac{1}{q^{sj}}\right)\right)
\end{align*}
and the theorem statement follows.
\end{IEEEproof}

\section{Parameter Choices} \label{sec:parameters}

\begin{table}\centering
\caption{Parameter choices and the work factors of the attacks presented in Section~\ref{sec:subspaceAttack} and \ref{sec:linDepAttack}. The rate $R$ is obtained by Corollary~\ref{cor:PIRrate}.}
\label{fig:parameters}
\begin{tabular}{CCCCCCC|CC}
  q&s&v&n&k&\delta&R& \ref{sec:subspaceAttack} & \ref{sec:linDepAttack} \\ \hline
  16&32&31&100&50&50&\frac{1}{64}&2^{124}&2^{128} \\[1ex]
  16 & 32 & 16 & 100 & 50 & 800 & \frac{1}{4} & 2^{64} & 2^{128} \\[1ex]
  32&32&31&100&50&50&\frac{1}{64}&2^{155}&2^{160} \\[1ex]
  32&32&26&100&50&50&\frac{3}{32}&2^{130}&2^{160} \\[1ex]
  32& 32& 24 & 100 & 50 & 400 & \frac{1}{8}& 2^{160} & 2^{120}\\[1ex]
  64&32&21&100&50&550&\frac{11}{64}&2^{192}&2^{126}
\end{tabular}
\end{table}

\begin{figure}
    \centering
    \begin{tikzpicture}
\pgfplotsset{compat = 1.3}
\begin{axis}[
	legend style={nodes={scale=0.7, transform shape}},
	width = 0.89\columnwidth,
	xlabel = $v$,
	xlabel style = {nodes={scale=0.85, transform shape}},
	ylabel = Complexity,
	ylabel style={nodes={scale=0.85, transform shape}},
    ymode = log,
    log basis y={2},
	xmin = 10,
	xmax = 32,
	legend pos = south east]

\addplot[color=blue, mark=x,smooth] table {SSBoundFile.txt};
\addlegendentry{\ref{sec:subspaceAttack}}
\addplot[color=red, mark=o,smooth] table {LDBoundFile.txt};
\addlegendentry{\ref{sec:linDepAttack}}

\end{axis}
\end{tikzpicture}

    \caption{Attack complexities for different values of $v$ with $q=32$, $s=32$, $n=100$, and $k=50$.}
    \label{fig:my_label}
\end{figure}
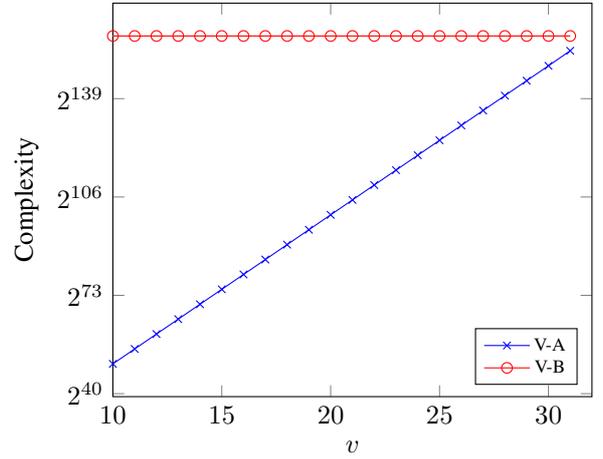

Table~\ref{fig:parameters} shows the achieved PIR rate for different choices of parameters together with lower bounds on the complexity of the respective attacks, as derived in Section~\ref{sec:subspaceAttack} and~\ref{sec:linDepAttack}. Note that rate of the presented scheme depends greatly on the chosen parameters. Increasing $q$ and/or $s$ increases the security and therefore allows for increasing the rate of the scheme by adapting $v$ and/or $k$. However, increasing the values of $q$ or $s$ increases the complexity of the scheme, as the server is required to perform multiplications over the respective fields. As the computational complexity is regarded as the bottleneck for computational PIR \cite{sion2007computational}, we present parameters resulting in a low rate, but relatively good complexity.
Due to a lack of space a detailed comparison of the complexity compared to the existing schemes of \cite{aguilar2007lattice,yi2012single} is left as future work. Instead we provide some intuition on why the scheme can perform favorably compared to these schemes in terms of complexity.
Although the field size resulting from the parameters given in Table~\ref{fig:parameters} appear to be large from a coding-theoretic point of view, the majority of the more complex operations, i.e., multiplications, is not over these fields, but instead between elements of the field and elements of a subfield. Especially, since the files are only from $\Fq$, all multiplications performed on the server side, the number of which depends on the (generally large) number of files \emph{and} their size, are of the form $\alpha \beta$ with $\alpha \in \Fqs$ and $\beta \in \Fq$. Each element $\alpha \in \Fqs$ can be represented as a polynomial of degree $s-1$ over $\Fq$, so the complexity of this multiplication is just the complexity of multiplying the $s$ coefficients of this polynomial by $\beta$. Assuming a complexity of $(\log(q))^2$ for the multiplication of elements from a field $\Fq$, this gives a complexity of $s (\log(q))^2 = (\log(q^{\sqrt{s}}))^2 $, which is equivalent to performing multiplications over a field $\F_{q^{\sqrt{s}}}$. For example, for $q=32$ and $s=32$ this is approximately equivalent to the complexity of multiplications over $\F_{2^{29}}$. 
As a comparison, the parameters proposed in \cite[Section~IV]{aguilar2007lattice} require the multiplication of matrices of similar size to our scheme on the server side, but over the integer field $\F_{2^{60}+325}$. This is not only significantly larger than the "equivalent field" in our construction, but additionally does not provide the hardware advantages that extension fields of $\F_{2}$ provide, namely the possibility of implementation based on shifts and~XORs. 

\todoLater{\aw{We should also say something about the rates compared to [9]. Or do we only have this factor 6 as a rate??}

\lh{We ignore the query generation here. There we actually have to do calculations over $\F_{q^s}$ and the number of such multiplications depends on the number of files. This is justified if we assume the files to be very large, since the query size is independent of that.}}

\bibliographystyle{IEEEtran}
\bibliography{main,IT-PIR}

\begin{thebibliography}{10}
\providecommand{\url}[1]{#1}
\csname url@samestyle\endcsname
\providecommand{\newblock}{\relax}
\providecommand{\bibinfo}[2]{#2}
\providecommand{\BIBentrySTDinterwordspacing}{\spaceskip=0pt\relax}
\providecommand{\BIBentryALTinterwordstretchfactor}{4}
\providecommand{\BIBentryALTinterwordspacing}{\spaceskip=\fontdimen2\font plus
\BIBentryALTinterwordstretchfactor\fontdimen3\font minus
  \fontdimen4\font\relax}
\providecommand{\BIBforeignlanguage}[2]{{%
\expandafter\ifx\csname l@#1\endcsname\relax
\typeout{** WARNING: IEEEtran.bst: No hyphenation pattern has been}%
\typeout{** loaded for the language `#1'. Using the pattern for}%
\typeout{** the default language instead.}%
\else
\language=\csname l@#1\endcsname
\fi
#2}}
\providecommand{\BIBdecl}{\relax}
\BIBdecl

\bibitem{chor1995private}
B.~Chor, O.~Goldreich, E.~Kushilevitz, and M.~Sudan, ``Private information
  retrieval,'' in \emph{Proceedings of IEEE 36th Annual Foundations of Computer
  Science}.\hskip 1em plus 0.5em minus 0.4em\relax IEEE, 1995, pp. 41--50.

\bibitem{dvir20162}
Z.~Dvir and S.~Gopi, ``2-server pir with subpolynomial communication,''
  \emph{Journal of the ACM (JACM)}, vol.~63, no.~4, p.~39, 2016.

\bibitem{beimel2002breaking}
A.~Beimel, Y.~Ishai, E.~Kushilevitz, and J.-F. Raymond, ``Breaking the {O}
  (n/sup 1/(2k-1)/) barrier for information-theoretic private information
  retrieval,'' in \emph{The 43rd Annual IEEE Symposium on Foundations of
  Computer Science, 2002. Proceedings.}\hskip 1em plus 0.5em minus 0.4em\relax
  IEEE, 2002, pp. 261--270.

\bibitem{sun2018capacity}
H.~Sun and S.~A. Jafar, ``The capacity of symmetric private information
  retrieval,'' \emph{IEEE Transactions on Information Theory}, vol.~65, no.~1,
  pp. 322--329, 2018.

\bibitem{sun2017capacity}
------, ``The capacity of robust private information retrieval with colluding
  databases,'' \emph{IEEE Transactions on Information Theory}, vol.~64, no.~4,
  pp. 2361--2370, 2017.

\bibitem{banawan2018capacity}
K.~Banawan and S.~Ulukus, ``The capacity of private information retrieval from
  coded databases,'' \emph{IEEE Transactions on Information Theory}, vol.~64,
  no.~3, pp. 1945--1956, 2018.

\bibitem{freij2017private}
R.~Freij-Hollanti, O.~W. Gnilke, C.~Hollanti, and D.~A. Karpuk, ``Private
  information retrieval from coded databases with colluding servers,''
  \emph{SIAM Journal on Applied Algebra and Geometry}, vol.~1, no.~1, pp.
  647--664, 2017.

\bibitem{kushilevitz1997replication}
E.~Kushilevitz and R.~Ostrovsky, ``Replication is not needed: Single database,
  computationally-private information retrieval,'' in \emph{Proceedings 38th
  Annual Symposium on Foundations of Computer Science}.\hskip 1em plus 0.5em
  minus 0.4em\relax IEEE, 1997, pp. 364--373.

\bibitem{lipmaa2005oblivious}
H.~Lipmaa, ``An oblivious transfer protocol with log-squared communication,''
  in \emph{International Conference on Information Security}.\hskip 1em plus
  0.5em minus 0.4em\relax Springer, 2005, pp. 314--328.

\bibitem{gentry2005single}
C.~Gentry and Z.~Ramzan, ``Single-database private information retrieval with
  constant communication rate,'' in \emph{International Colloquium on Automata,
  Languages, and Programming}.\hskip 1em plus 0.5em minus 0.4em\relax Springer,
  2005, pp. 803--815.

\bibitem{sion2007computational}
R.~Sion and B.~Carbunar, ``On the computational practicality of private
  information retrieval,'' in \emph{Proceedings of the Network and Distributed
  Systems Security Symposium}.\hskip 1em plus 0.5em minus 0.4em\relax Internet
  Society, 2007, pp. 2006--06.

\bibitem{aguilar2007lattice}
C.~Aguilar-Melchor and P.~Gaborit, ``A lattice-based computationally-efficient
  private information retrieval protocol,'' \emph{Cryptol. ePrint Arch.,
  Report}, vol. 446, 2007.

\bibitem{liu2016cryptanalysis}
J.~Liu and J.~Bi, ``Cryptanalysis of a fast private information retrieval
  protocol,'' in \emph{Proceedings of the 3rd ACM International Workshop on
  ASIA Public-Key Cryptography}.\hskip 1em plus 0.5em minus 0.4em\relax ACM,
  2016, pp. 56--60.

\bibitem{gentry2009fully}
C.~Gentry and D.~Boneh, \emph{A fully homomorphic encryption scheme}.\hskip 1em
  plus 0.5em minus 0.4em\relax Stanford University Stanford, 2009, vol.~20,
  no.~09.

\bibitem{yi2012single}
X.~Yi, M.~G. Kaosar, R.~Paulet, and E.~Bertino, ``Single-database private
  information retrieval from fully homomorphic encryption,'' \emph{IEEE
  Transactions on Knowledge and Data Engineering}, vol.~25, no.~5, pp.
  1125--1134, 2012.

\bibitem{kiayias2015optimal}
A.~Kiayias, N.~Leonardos, H.~Lipmaa, K.~Pavlyk, and Q.~Tang, ``Optimal rate
  private information retrieval from homomorphic encryption,''
  \emph{Proceedings on Privacy Enhancing Technologies}, vol. 2015, no.~2, pp.
  222--243, 2015.

\bibitem{aguilar2016xpir}
C.~Aguilar-Melchor, J.~Barrier, L.~Fousse, and M.-O. Killijian, ``{XPIR}:
  Private information retrieval for everyone,'' \emph{Proceedings on Privacy
  Enhancing Technologies}, vol. 2016, no.~2, pp. 155--174, 2016.

\bibitem{lipmaa2017simpler}
H.~Lipmaa and K.~Pavlyk, ``A simpler rate-optimal {CPIR} protocol,'' in
  \emph{International Conference on Financial Cryptography and Data
  Security}.\hskip 1em plus 0.5em minus 0.4em\relax Springer, 2017, pp.
  621--638.

\bibitem{gentry2019compressible}
C.~Gentry and S.~Halevi, ``Compressible {FHE} with applications to {PIR},'' in
  \emph{Theory of Cryptography Conference}.\hskip 1em plus 0.5em minus
  0.4em\relax Springer, 2019, pp. 438--464.

\bibitem{angel2018pir}
S.~Angel, H.~Chen, K.~Laine, and S.~Setty, ``{PIR} with compressed queries and
  amortized query processing,'' in \emph{2018 IEEE Symposium on Security and
  Privacy (SP)}.\hskip 1em plus 0.5em minus 0.4em\relax IEEE, 2018, pp.
  962--979.

\bibitem{bordage2020privacy}
S.~Bordage and J.~Lavauzelle, ``On the privacy of a code-based single-server
  computational pir scheme,'' 2020.

\end{thebibliography}

\end{document}